\documentclass{amsart}

\usepackage[foot]{amsaddr}                  
\usepackage{amsmath}                        
\usepackage{amssymb}                        
\usepackage{dsfont} 	                    
\usepackage{bm}                             
\usepackage{mathtools}                      
\usepackage[hypertexnames=false]{hyperref}  
\usepackage[scientific-notation=true]{siunitx}
\usepackage{blkarray}
\usepackage{enumerate}
\usepackage{centernot}
\usepackage{paralist}
\usepackage{stackrel}
\usepackage[table]{xcolor}
\usepackage[colorinlistoftodos,prependcaption,textsize=tiny]{todonotes} 
\usepackage{pagenote}
\usepackage{setspace}
\usepackage[english]{babel}
\usepackage{subfig}
\usepackage{graphicx}
\usepackage{subfig}
\usepackage{dirtytalk}

\usepackage[authoryear]{natbib}

\usepackage{algorithm}
\usepackage{algpseudocode}

\usepackage{listings}
\usepackage{color}

\usepackage{amsthm}
\theoremstyle{plain}
\theoremstyle{definition}

\theoremstyle{remark}

\newcommand{\LA}{\mathbf{\Lambda}}

\newcommand{\PS}{\mathbf{\Psi}}

\newcommand{\Covmat}{\boldsymbol{\Sigma}}

\newcommand{\G}{\mathbf{G}}

\newcommand{\K}{\mathbf{K}}

\newcommand{\C}{\mathbf{C}}
\newcommand{\D}{\mathbf{D}}
\newcommand{\s}{\mathbf{s}}

\newcommand{\V}{\mathbf{V}}

\newcommand{\I}{\mathbf{I}}
\newcommand{\scores}{\mathbf{s}}

\newcommand{\update}{\boldsymbol{\delta}}
\newcommand{\mstrut}{\rule{0pt}{2.2ex}}
\newcommand{\y}{\mathbf{y}}
\newcommand{\X}{\mathbf{X}}
\newcommand{\blues}{\boldsymbol{\beta}}
\newcommand{\Z}{\mathbf{Z}}
\newcommand{\blups}{\mathbf{u}}
\newcommand{\solution}{\boldsymbol{\tau}}
\newcommand{\e}{\boldsymbol{\epsilon}}
\newcommand{\Gparams}{\boldsymbol{\theta}}
\newcommand{\Rparams}{\boldsymbol{\phi}}
\newcommand{\params}{\boldsymbol{\kappa}}
\newcommand{\R}{\mathbf{R}}
\newcommand{\Q}{\mathbf{Q}}
\newcommand{\Pmat}{\mathbf{P}}
\newcommand{\Vmat}{\mathbf{V}}
\newcommand{\Wmat}{\mathbf{W}}
\newcommand{\AImat}{\boldsymbol{\mathcal{I}}}

\newcommand{\WV}{\boldsymbol{\mathcal{Y}}}
\newcommand{\wv}{\boldsymbol{y}}

\newcommand{\LT}{\mathbf{L}}

\makeatletter
\def\squarecorner#1{
	\pgf@x=\the\wd\pgfnodeparttextbox%
	\pgfmathsetlength\pgf@xc{\pgfkeysvalueof{/pgf/inner xsep}}%
	\advance\pgf@x by 2\pgf@xc%
	\pgfmathsetlength\pgf@xb{\pgfkeysvalueof{/pgf/minimum width}}%
	\ifdim\pgf@x<\pgf@xb%
	\pgf@x=\pgf@xb%
	\fi%
	\pgf@y=\ht\pgfnodeparttextbox%
	\advance\pgf@y by\dp\pgfnodeparttextbox%
	\pgfmathsetlength\pgf@yc{\pgfkeysvalueof{/pgf/inner ysep}}%
	\advance\pgf@y by 2\pgf@yc%
	\pgfmathsetlength\pgf@yb{\pgfkeysvalueof{/pgf/minimum height}}%
	\ifdim\pgf@y<\pgf@yb%
	\pgf@y=\pgf@yb%
	\fi%
	\ifdim\pgf@x<\pgf@y%
	\pgf@x=\pgf@y%
	\else
	\pgf@y=\pgf@x%
	\fi
	\pgf@x=#1.5\pgf@x%
	\advance\pgf@x by.5\wd\pgfnodeparttextbox%
	\pgfmathsetlength\pgf@xa{\pgfkeysvalueof{/pgf/outer xsep}}%
	\advance\pgf@x by#1\pgf@xa%
	\pgf@y=#1.5\pgf@y%
	\advance\pgf@y by-.5\dp\pgfnodeparttextbox%
	\advance\pgf@y by.5\ht\pgfnodeparttextbox%
	\pgfmathsetlength\pgf@ya{\pgfkeysvalueof{/pgf/outer ysep}}%
	\advance\pgf@y by#1\pgf@ya%
}
\makeatother

\pgfdeclareshape{square}{
	\savedanchor\northeast{\squarecorner{}}
	\savedanchor\southwest{\squarecorner{-}}
	
	\foreach \x in {east,west} \foreach \y in {north,mid,base,south} {
		\inheritanchor[from=rectangle]{\y\space\x}
	}
	\foreach \x in {east,west,north,mid,base,south,center,text} {
		\inheritanchor[from=rectangle]{\x}
	}
	\inheritanchorborder[from=rectangle]
	\inheritbackgroundpath[from=rectangle]
}

\def\ImportFromMnSymbol#1{%
	\DeclareFontFamily{U} {MnSymbol#1}{}
	\DeclareFontShape{U}{MnSymbol#1}{m}{n}{
		<-6> MnSymbol#15
		<6-7> MnSymbol#16
		<7-8> MnSymbol#17
		<8-9> MnSymbol#18
		<9-10> MnSymbol#19
		<10-12> MnSymbol#110
		<12-> MnSymbol#112}{}
	\DeclareFontShape{U}{MnSymbol#1}{b}{n}{
		<-6> MnSymbol#1-Bold5
		<6-7> MnSymbol#1-Bold6
		<7-8> MnSymbol#1-Bold7
		<8-9> MnSymbol#1-Bold8
		<9-10> MnSymbol#1-Bold9
		<10-12> MnSymbol#1-Bold10
		<12-> MnSymbol#1-Bold12}{}
	\DeclareSymbolFont{MnSy#1} {U} {MnSymbol#1}{m}{n}
}
\newcommand\DeclareMnSymbol[4]{\DeclareMathSymbol{#1}{#2}{MnSy#3}{#4}}
\ImportFromMnSymbol{A}
\DeclareMnSymbol{\ConIndepNat}{\mathrel}{A}{225}

\makeatletter
\newcommand{\addresseshere}{%
  \enddoc@text\let\enddoc@text\relax
}

\definecolor{lightgrey}{rgb}{0.9,0.9,0.9}
\definecolor{darkgreen}{rgb}{0,0.6,0}
\calclayout

\usepackage{enumitem}
\usepackage{tikz}
\tikzstyle{node}=[very thick, circle, draw=black, minimum size=22, inner sep=0.8, outer sep=0.6]

\usetikzlibrary{calc}
\usetikzlibrary{arrows.meta}
\usetikzlibrary{chains}
\usetikzlibrary{arrows, arrows.meta}

\DeclareRobustCommand{\rvdots}{%
	\vbox{
		\baselineskip4\p@\lineskiplimit\z@
		\kern-\p@
		\hbox{.}\hbox{.}\hbox{.}
}}

\begin{document}
\sloppy

\title[Hybrid GPU/CPU average information REML]
{A hybrid GPU/CPU implementation of the average information restricted maximum likelihood algorithm for fitting multivariate linear mixed models}

\author{Killian A.C.\ Melsen$^{1*}$}
\author{Fred A.\ van Eeuwijk$^1$}
\author{Carel F.W.\ Peeters$^1$}

\address{$^1$Mathematical \& Statistical Methods group - Biometris, Wageningen University \& Research, PO Box 16, 6700 AA, Wageningen, The Netherlands}

\email{$^*$killian.melsen@wur.nl}

\begin{abstract}
\label{abstract}
The multivariate linear mixed model combined with restricted maximum likelihood estimation of (co)variance components and solving the mixed model equations is widely used for the analysis of plant breeding data.
Several algorithms for restricted maximum likelihood estimation have been proposed, with the average information algorithm being widely used for its computational efficiency, quadratic rate of convergence, and ability to handle large amounts of parameters.
A key reason for the scalability of linear mixed models is the exploitation of sparse matrix techniques.
Recently, decreasing costs of high-throughput data collection have led to high-dimensional sets of features characterizing genotypes and environments.
A convenient way of using these features is through the use of kernel-based covariance structures.
However, while convenient, these kernels typically have dense inverses meaning optimizations based on sparse matrix techniques are not always applicable.
For these settings, the introduction of general-purpose computing on graphics processing units brought significant computational advantages for dense linear algebra operations.

We present a hybrid GPU/CPU implementation of the average information restricted maximum likelihood algorithm.
We eliminate sparse equations using supernodal Cholesky factorization and rely on CUDA GPU acceleration to solve the resulting reduced dense system.
We show the significant computational advantages of this approach for fully dense problems, as well as problems with a mix of dense and sparse equations.

\end{abstract}

\maketitle


\section{Introduction}\label{SEC:Intro}
Linear mixed models combined with restricted maximum likelihood (REML) estimation of (co)variance components have been one of the most widely used statistical tools in the analysis of plant breeding data \citep{pattersonandthompson_1971, Henderson_1953}.
Several algorithms for fitting these models have been proposed, starting with Fisher scoring as used in the original version of REML by \cite{pattersonandthompson_1971}.
Computational limitations then led to the development of derivative-free and expectation maximization approaches, and finally the average information (AI) algorithm which avoided calculation of computationally prohibitive traces \citep{Thompson_2019, Gilmouretal_1995, johnsonandthompson_1995}.
For a review of the different algorithms we refer to \citep{Hofer_1998}.

Linear mixed models based on AI REML quickly found several applications, ranging from the modeling of spatial trends to multivariate analyses \citep{Jensenetal_1997}.
Several authors have reviewed its applications, including \citet{Gilmour_2019} who dealt with the implementation in the widely used software packages {\ttfamily ASReml}, {\ttfamily ASReml-R}, and recently {\ttfamily ASReml-P} \citep{VSNi_2026, butleretal_2023}.

A key reason for the scalability of linear mixed models in general is the application of sparse matrix techniques.
A significant amount of work has gone into optimizing algorithms for solving sparse linear systems like the ones encountered in linear mixed models \citep{Mazur_2023, GilmourandThompson_2006, Meyer_2005, Amestoyetal_2001, GeorgeandLiu_1989}.
Note however that this issue is not at all restricted to the field of genetics and linear mixed models \citep{Takahashietal_1973}.
Others have focused on developing sparse implementations of existing models.
For example, \citet{Thompsonetal_2003} introduced a sparse implementation of the commonly used factor analytic covariance structure for the analysis of genotype by environment (G$\times$E) interactions \citep{tolhurstetal_2019}.
Another example is the sparse implementation of splines for modeling spatial trends in field trials \citep{Boer_2023}.

With the cost of high-throughput pheno-, geno-, and envirotyping going down, high-dimensional datasets characterizing genotypes and environments have become increasingly common.
These data can be incorporated in linear mixed models in several ways.
Single nucleotide polymorphism (SNP) marker information is often included through kinship or genomic relationship matrices (GRMs), while environmental covariables can be conveniently included through reaction norm models or linear and nonlinear kernels \citep{Melsenetal_2026, Costa-Netoetal_2021, Jarquinetal_2014, VanRaden_2008}.
These approaches lower the amount of parameters to be estimated compared to unstructured or even factor analytic models while complex interactions are still included.

However, from a computational perspective these developments change the structure of the mixed model equations (MMEs).
Traditional pedigree-based relationship matrices for genotypes typically have a sparse inverse, and when combined with, e.g., the extended factor analytic structure and spatial effects, result in a highly sparse MME coefficient matrix.
On the other hand, GRMs and many other kernels typically have dense inverses.
As a result, modern linear mixed models may contain a coefficient matrix that has significant dense blocks.
Treating the complete mixed model equations as sparse means modern parallelized solvers are not fully exploited.
Treating all equations as dense, on the other hand, means specialized approaches for sparse matrices cannot be used.

Recent developments in high-performance computing are especially relevant to the mix of dense and sparse structures in the MMEs.
Graphics processing units (GPUs) provide very high throughput for dense linear algebra operations such as matrix multiplication and dense Cholesky factorization \citep{Tomovetal_2010, Luebke2008, VolkovandDemmel_2008}.
While efficient for dense operations, their advantages are generally less obvious for, e.g., sparse factorizations.
Supernodal sparse Cholesky factorization provides a natural interface between these two frameworks by representing sets of columns of a sparse Cholesky factor as dense
matrices, and has been applied to the fitting of linear mixed models \citep{Mazur_2023, Masudaetal_2015, Masudaetal_2014}.
Given the respective benefits of both dense and sparse computational frameworks, modern mixed models may benefit from a hybrid GPU/CPU implementation instead of an exclusively sparse or dense approach.

Some linear mixed model software has recently included GPU acceleration \citep{MarchalandRaimondi_2026, Rodriguezetal_2026}.
These implementations show the usefulness of GPUs, but currently involve a significant trade-off between computational efficiency and wide applicability.
Implementations developed for large-scale genomic association analyses, for example, do not provide the flexibility in terms of specifying covariance matrices required for the analysis of G$\times$E data.
Furthermore, implementations based on fully dense linear algebra can make effective use of GPUs but, as mentioned above, do not exploit sparsity.
General-purpose AI REML software used in plant breeding therefore remains mostly based on CPU implementations of sparse linear algebra.

We develop a hybrid GPU/CPU implementation of the AI REML algorithm designed specifically for linear mixed models containing both dense and sparse structures.
Instead of choosing between either a fully sparse or fully dense dense approach to solving the MMEs, we eliminate the sparse part of the system using a supernodal Cholesky factorization and solve the dense Schur complement on the GPU.
This retains flexibility in terms of exploiting both dense and sparse computational frameworks and includes support for unstructured, extended factor analytic, autoregressive, and kernel-based covariance structures.

The goal of this paper is three-fold.
We first review the general theory of the AI REML approach to fitting multivariate linear mixed models in Section \ref{sec:theory}.
We then develop a hybrid GPU/CPU implementation based on supernodal Cholesky factorization and a dense Schur complement in Section \ref{sec:algo}.
Finally, we investigate some extensions of the algorithm and evaluate its computational performance for a few commonly used linear mixed models containing different combinations of dense and sparse covariance structures in Sections \ref{sec:extensions} and \ref{sec:illustrations}.
A discussion of potential further work is provided in Section \ref{sec:discussion}.

\section{Average information REML}\label{sec:theory}
Consider the linear mixed model
\begin{equation}\label{eq:LMM}
	\y = \X\blues + \Z\blups + \e \text{,}
\end{equation}
where $\y$ is the vector of phenotypes, $\X$ is a fixed effects design matrix, $\blues$ is the vector containing best linear unbiased estimates (BLUEs) for the fixed effects, $\Z$ is a random effects design matrix, $\blups$ is the vector of best linear unbiased predictions (BLUPs) for the random effects, and $\e$ contains residuals.
Using notation from \citet{Gilmour_2019}, the BLUPs are assumed to be normally distributed with mean vector $\mathbf{0}$ and some covariance matrix $\G$.
Similarly, $\e \sim \mathcal{N}\left(\mathbf{0}, \R \right)$.
Note that $\G$ is a function of parameters in $\Gparams$ and $\R$ is a function of parameters in $\Rparams$ so that the full set of parameters $\params = [\Gparams^\top, \Rparams^\top]^\top$ where $\Gparams^\top = \left[\theta_1\dots\theta_a\dots\theta_g\right]$ and $\Rparams^\top = \left[\phi_1\dots\phi_b\dots\phi_r\right]$.

To model $p$, independent random effects, $\G$ can be a direct sum of smaller covariance matrices $\G_i$:
\begin{equation*}
	\begin{split}
		\G &= \oplus_{i = 1}^p \G_i = \begin{bmatrix}
			\G_1 & 0 & \dots & 0 \\
			0 & \G_2 & \dots & 0 \\
			\vdots & \vdots & \ddots & \vdots \\
			0 & 0 & \dots & \G_p
		\end{bmatrix} \text{,}
	\end{split}
\end{equation*}
where each individual $\G_i$ is in turn allowed to be a Kronecker product of covariance matrices associated with different factors in the data.
For modeling G$\times$E interactions in plant breeding, for example, $\G_i$ could be structured as $\Covmat_i \otimes \K_i$ where $\Covmat_i\in\mathbb{R}^{\eta_i \times \eta_i}$ is some covariance matrix for environments to be estimated from the data, and $\K_i\in\mathbb{R}^{f_i \times f_i}$ is a known additive genomic relationship or kinship matrix.
Indeed, each $\G_i$ may be written as a Kronecker product of some variable structure $\Covmat_i$ to be estimated and a non-variable, known structure $\K_i$.
Note that $\Covmat_i$ and $\K_i$ themselves may be Kronecker products when modeling complex higher-order interactions \citep{Melsenetal_2026}.

To fit model (\ref{eq:LMM}) we first need estimates for $\params$.
We can then use the MMEs to obtain estimates for $\solution = \left[\blues^\top,\blups^\top\right]^\top$:
\begin{equation}\label{eq:MME}
	\begin{bmatrix*}
		\X^\top \R^{-1} \X & \X^\top \R^{-1} \Z \\
		\Z^\top \R^{-1} \X & \Z^\top \R^{-1} \Z + \G^{-1}
	\end{bmatrix*}
	\begin{bmatrix*}
		\blues \\
		\blups
	\end{bmatrix*} =
	\begin{bmatrix*}
		\X^\top \R^{-1} \y \\
		\Z^\top \R^{-1} \y
	\end{bmatrix*} \text{.}
\end{equation}
To obtain estimates for $\params$, we maximize the REML log-likelihood using the AI algorithm.
The REML log-likelihood, ignoring a constant, is as follows:
\begin{equation}\label{eq:LL}
	l = -\dfrac{1}{2} \left(\log|\C| + \log|\R| + \log|\G| + \y^\top \Pmat \y\right) \text{,}
\end{equation}
where $\C$ is the coefficient matrix on the left side of (\ref{eq:MME}) and $\Pmat = \Vmat^{-1} - \Vmat^{-1} \X (\X^\top \Vmat^{-1} \X)^{-1} \X^\top \Vmat$, where $\Vmat = \Z\G\Z^\top + \R$.

To compute updates for all parameters, given a vector of current parameter values $\params_t$, we need the matrix of negative second derivatives of $l$ with respect to $\params$, as well as the vector of first derivatives.
The first is referred to as the AI matrix $\AImat$, while the second is typically referred to as the vector of scores $\scores$.
The AI matrix can be obtained as $\AImat = \WV^\top \Pmat \WV / 2$ where $\WV = \left[\WV_{\Gparams}\;\WV_{\Rparams}\right]$ is the matrix of working variates for parameters involved in the random effects and residual structure.
Note that $\WV_{\Gparams} = \left[\wv_{1}\dots\wv_{a}\dots\wv_{g}\right]$ and $\WV_{\Rparams} = \left[\wv_{1}\dots\wv_{b}\dots\wv_{r}\right]$.
Each random effect working variate $\wv_a$ corresponding to parameter $\theta_a$ can be obtained as:
\begin{equation}\label{eq:randomwv}
	\wv_a = \Z\dfrac{\partial\G}{\partial\theta_a}\G^{-1}\hat{\blups} \text{,}
\end{equation}
where $\hat{\blups}$ (as well as $\hat{\blues}$) is obtained by solving (\ref{eq:MME}) given the current parameters $\params$.
Similarly, the working variates for parameters involved in the residual structure $\wv_b$ can be obtained as:
\begin{equation*}
	\wv_b = \dfrac{\partial\R}{\partial\phi_b}\R^{-1}\hat{\e} \text{,}
\end{equation*}
where $\hat{\e} = \y - \X\hat{\blues} - \Z\hat{\blups}$.

Like the working variates, the vector of scores can be divided into scores related to random effects and scores related to the residual structure.
Each score $s_a$ for a parameter in $\Gparams$ can be obtained as:
\begin{equation*}
	s_a = -\dfrac{1}{2}\left[\text{tr}\left(\dfrac{\partial\G}{\partial\theta_a}\G^{-1}\right) - \text{tr}\left(\G^{-1}\dfrac{\partial\G}{\partial\theta_a}\G^{-1}\C_{-zz}\right) - \y^\top\Pmat\wv_a\right] \text{,}
\end{equation*}
where the matrix $\C_{-zz}$ corresponds to the lower-right block of $\C^{-1}$.
Similarly, for the residual effect:
\begin{equation}\label{eq:naiveresscores}
	s_b = -\dfrac{1}{2}\left[\text{tr}\left(\dfrac{\partial\R}{\partial\phi_b}\R^{-1}\right) - \text{tr}\left(\Wmat^\top \R^{-1} \dfrac{\partial\R}{\partial\phi_b} \R^{-1} \Wmat\C^{-1}\right) - \y^\top\Pmat\wv_b\right] \text{,}
\end{equation}
where $\Wmat = \left[\X\;\Z\right]$.

Having obtained the AI matrix and vector of scores, Newton-Raphson updates for the parameters can be computed as:
\begin{equation*}
	\params_{t + 1} = \params_t + \AImat^{-1} \scores \text{.}
\end{equation*}

\section{Algorithm}\label{sec:algo}
From a computational perspective, the main challenges involved in fitting the linear mixed model using the MMEs are the inversion of the coefficient matrix $\C$ and obtaining $\Pmat$, which requires inverting the phenotypic covariance matrix $\Vmat$.
Working with $\Pmat$ directly can be avoided throughout the algorithm and inversion of $\C$ can be done efficiently using GPU acceleration for the dense part and the Takahashi equations for the sparse part \citep{Luebke2008, Takahashietal_1973}.
The remainder of this section provides details on our hybrid implementation of the AI REML algorithm.

\subsection{Partitioning of the equations}
The first step of the iterative algorithm involves constructing the residual covariance matrix $\R = \Covmat_\epsilon \otimes \K_\epsilon$, where , e.g., $\Covmat_\epsilon$ is an unstructured covariance matrix of traits to be estimated and $\K_\epsilon = \I$ for plots.
The full matrix $\R$ can then be subsetted to match the non-missing data and inverted via a Cholesky factorization to cheaply obtain $\text{log}|\R|$ as well.
As both $\R$ and $\R^{-1}$ are typically diagonal, block-diagonal, or banded, and thus highly sparse, Cholesky factorization and inversion are computationally negligible.

The base coefficient matrix $\C^*$ can be formed after obtaining $\R^{-1}$.
From a theoretical perspective, a simple partitioning of $\C^*$ into blocks related to fixed and random effects as shown in (\ref{eq:MME}) is convenient.
From a computational viewpoint, however, $\C^*$ needs to be partitioned into blocks related to dense and sparse sets of equations first:
\begin{equation*}
	\C^* = \begin{bmatrix*}
		\C^*_{dd} & \C^*_{ds} \\
		\C^*_{sd} & \C^*_{ss}
	\end{bmatrix*} \text{.}
\end{equation*}
Note that a fixed effect equation can be deemed dense if the proportion of non-zeros in the relevant column of $\X$ exceeds some threshold.
Alternatively, all fixed effect equations can simply be treated as dense for simplicity, or sparse if an overall intercept and covariates are missing.
Similarly, the set of equations for a given random effect $i$ are deemed dense if the Kronecker product to form $\G_i$ involves dense structures only.
Each block of the base coefficient matrix can now be formed:
\begin{equation*}
	\begin{split}
		\C^*_{dd} &= \Wmat_{d}^\top \R^{-1} \Wmat_{d} \\
		\C^*_{ds} &= \C^{*\top}_{sd} = \Wmat_{d}^\top \R^{-1} \Wmat_{s} \\
		\C^*_{ss} &= \Wmat_{s}^\top \R^{-1} \Wmat_{s}
	\end{split} \text{,}
\end{equation*}
where $\Wmat_{d} = \left[\X_{d}\;\Z_{d}\right]$ and $\Wmat_{s} = \left[\X_{s}\;\Z_{s}\right]$ are constructed using the columns of the design matrices related to dense and sparse equations, respectively.
Note that each block of $\C^*$ can thus be partitioned into sub-blocks similar to (\ref{eq:MME}), e.g.:
\begin{equation*}
	\C^*_{ss} = \begin{bmatrix*}
		\X_s^\top \R^{-1} \X_s & \X_s^\top \R^{-1} \Z_s \\
		\Z_s^\top \R^{-1} \X_s & \Z_s^\top \R^{-1} \Z_s
	\end{bmatrix*} \text{.}
\end{equation*}
Also note that in the special cases where either all equations are dense or sparse, no partitioning is required and the algorithm simplifies significantly.
For completeness, we will cover the case of a mix of dense and sparse equations in the remainder.

\subsection{Forming each $\G_i^{-1}$ and completing $\C$}
To complete the coefficient matrix, all inverse random effect covariance matrices $\G_i^{-1}$ need to be added to the appropriate diagonal blocks of $\C^*$.
Recall that each $\G_i = \Covmat_i \otimes \K_i$ and thus that each $\G_i^{-1} = \Covmat_i^{-1} \otimes \K_i^{-1}$.
As a result, we can form each inverse covariance matrix using the inverses of the component matrices, rather than naively inverting the full $\G_i$.
Note that all $\K_i^{-1}$ can be computed once before starting the iterative process as these matrices are known and do not need to be estimated.
Similarly, log determinants for all $\G_i$ can be efficiently obtained as a weighted linear combination of log determinants of the component matrices.
The weights are related to the dimensions of all components, and each $\text{log}|\K_i|$ can once again be computed before starting the iterative process.
Note that as the full $\G$ is strictly block-diagonal, $\text{log}|\G| = \sum_{i=1}^{p}\text{log}|\G_i|$.

After adding all inverse random effect covariance matrices to either $\C^*_{dd}$ or $\C^*_{ss}$ depending on whether the equations for random effect $i$ are dense or sparse, we obtain our full linear system $\C \solution = \Wmat^\top \R^{-1} \y$ partitioned into dense and sparse equations:
{\small
\begin{equation}\label{eq:MME_partitioned}
	\begin{bmatrix}
		\mstrut\X_{d}^\top \R^{-1} \X_{d} & \X_{d}^\top \R^{-1} \Z_{d}               & \X_{d}^\top \R^{-1} \X_{s} & \X_{d}^\top \R^{-1} \Z_{s} \\[3pt]
		\mstrut\Z_{d}^\top \R^{-1} \X_{d} & \Z_{d}^\top \R^{-1} \Z_{d} + \G_{d}^{-1} & \X_{d}^\top \R^{-1} \X_{s} & \X_{d}^\top \R^{-1} \Z_{s} \\[8pt]
		\mstrut\X_{s}^\top \R^{-1} \X_{d} & \Z_{d}^\top \R^{-1} \Z_{d}               & \X_{s}^\top \R^{-1} \X_{s} & \X_{s}^\top \R^{-1} \Z_{s} \\[3pt]
		\mstrut\X_{s}^\top \R^{-1} \X_{d} & \Z_{d}^\top \R^{-1} \Z_{d}               & \Z_{s}^\top \R^{-1} \X_{s} & \Z_{s}^\top \R^{-1} \Z_{s} + \G_{s}^{-1}
	\end{bmatrix}\begin{bmatrix}
		\mstrut\blues_{d} \\[3pt]
		\mstrut\blups_{d} \\[8pt]
		\mstrut\blues_{s} \\[3pt]
		\mstrut\blups_{s}
	\end{bmatrix} = \begin{bmatrix}
		\mstrut\X_{d}^\top \R^{-1} \y \\[3pt]
		\mstrut\Z_{d}^\top \R^{-1} \y \\[8pt]
		\mstrut\X_{s}^\top \R^{-1} \y \\[3pt]
		\mstrut\Z_{s}^\top \R^{-1} \y
	\end{bmatrix} \text{,}
\end{equation}
}
or, in more compact form:
\begin{equation*}
	\begin{bmatrix}
		\C_{dd} & \C_{ds} \\[4pt]
		\C_{sd} & \C_{ss}
	\end{bmatrix}\begin{bmatrix}
		\solution_d \\[4pt]
		\solution_s
	\end{bmatrix} = \begin{bmatrix}
		\Wmat_d^\top \R^{-1} \y \\[4pt]
		\Wmat_s^\top \R^{-1} \y
	\end{bmatrix} \text{.}
\end{equation*}

\subsection{Solving the linear system}\label{sec:solving}
To solve (\ref{eq:MME_partitioned}) we rely on supernodal Cholesky factorization of $\C_{ss}$ using {\ttfamily CHOLMOD} and its Schur complement $\C/\C_{ss} = \C_{dd} - \C_{ds} \C_{ss}^{-1} \C_{sd}$ \citep{Chenetal_2008}.
We first obtain the supernodal Cholesky factor $\LT_{ss}$ so that $\C_{ss} = \LT_{ss}\LT_{ss}^\top$ and compute $\text{log}|\C_{ss}| = 2\text{log}|\LT_{ss}|$.
Note that the log determinant of a Cholesky factor such as $\LT_{ss}$ can be obtained by simply summing the logarithms of the diagonal elements of $\LT_{ss}$.
We then solve the system:
\begin{equation*}
	\C_{ss}\Q = \C_{sd}
\end{equation*}
for $\Q$ using forward and backward substitution involving $\LT_{ss}$ and compute the Schur complement:
\begin{equation*}
	\C/\C_{ss} = \C_{dd} - \C_{ds} \Q\text{.}
\end{equation*}
We then, again using the Cholesky factor  $\LT_{ss}$, solve a second system:
\begin{equation*}
	\C_{ss}\boldsymbol{\gamma} = \Wmat_{s}^\top\R^{-1}\y
\end{equation*}
for the intermediate vector $\boldsymbol{\gamma}$ and compute the adjusted dense right-hand side $\boldsymbol{\xi}$ as:
\begin{equation*}
	\boldsymbol{\xi} = \Wmat_{d}^\top \R^{-1} \y - \C_{ds}\boldsymbol{\gamma}\text{.}
\end{equation*}
We then obtain the dense Cholesky factor of the Schur complement $\LT_{C/ss}$ and use it to compute $\text{log}|\C/\C_{ss}| = 2\text{log}|\LT_{C/ss}|$.
The log determinant of the full coefficient matrix $\text{log}|\C|$ is now easily obtained as $\text{log}|\C| = \text{log}|\C/\C_{ss}| + \text{log}|\C_{ss}|$.
Finally, the dense solutions are computed by solving the system:
\begin{equation*}
	\left(\C/\C_{ss}\right)\hat{\solution}_{d} = \boldsymbol{\xi}\text{,}
\end{equation*}
using $\LT_{C/ss}$, followed by computing the sparse solutions:
\begin{equation*}
	\hat{\solution}_{s} = \boldsymbol{\gamma} - \Q \hat{\solution}_{d}\text{.}
\end{equation*}
Some more details on the process outlined above are provided in Section \ref{SM:solving} of the Supplementary Material (SM).
Note that any systems involving $\C_{ss}$ can be solved efficiently by exploiting the sparsity in $\LT_{ss}$, and that for the dense system involving $\C/\C_{ss}$, the GPU or other parallel frameworks can be used to compute $\LT_{C/ss}$ and perform the forward and backward substitution.
A this point, the quadratic term $\y^\top \Pmat \y$ required for the log-likelihood can be computed as (see Sections \ref{SM:P} and \ref{SM:LL} of the SM):
\begin{equation}\label{eq:yPy}
	\y^\top \Pmat \y = \y^\top \R^{-1} \y - \y^\top \R^{-1} \Wmat_{d} \hat{\solution}_d - \y^\top \R^{-1} \Wmat_{s} \hat{\solution}_s
\end{equation}
followed by computing the log-likelihood using (\ref{eq:LL}).

\subsection{Working variates}
Recall that computing working variates for the parameters involved in each random effect $i$ using (\ref{eq:randomwv}) requires $\G_i^{-1}$, all derivatives for that random effect $\partial\G_i/\partial\theta_a$, and the BLUPs $\hat{\blups}_i$.
BLUPs and the inverse random effect covariance matrices are available at this point, and the derivatives are easily computed as Kronecker products using the smaller component matrices.
While we could naively compute the working variates using (\ref{eq:randomwv}), this would involve the full $\partial\G_i/\partial\theta_a$ and $\G_i^{-1}$, even if a large fixed structure $\K_i$ is involved.
Recall that the inverse covariance matrix for random effect $i$ can be written as:
\begin{equation*}
	\G_i^{-1} = \Covmat^{-1}_i \otimes \K_i^{-1} \text{.}
\end{equation*}
Similarly, note that
\begin{equation*}
	\dfrac{\partial\G_i}{\partial\theta_a} = \dfrac{\partial\Covmat_i}{\partial\theta_a} \otimes \K_i\text{.}
\end{equation*}
We can now rewrite (\ref{eq:randomwv}) using the mixed product property of the Kronecker product to avoid working with the full covariance matrices and derivatives:
\begin{equation}\label{eq:randomwvfast}
	\begin{split}
		\wv_a &= \Z_i\dfrac{\partial\G_i}{\partial\theta_a}\G_i^{-1}\hat{\blups}_i \\
		&= \Z_i\left(\dfrac{\partial\Covmat_i}{\partial\theta_a} \otimes \K_i\right)\left(\Covmat^{-1}_i \otimes \K_i^{-1}\right)\hat{\blups}_i \\
		&= \Z_i\left[\left(\dfrac{\partial\Covmat_i}{\partial\theta_a}\Covmat^{-1}_i\right)\otimes\left(\K_i\K_i^{-1}\right)\right]\hat{\blups}_i \\
		&= \Z_i\left[\left(\dfrac{\partial\Covmat_i}{\partial\theta_a}\Covmat^{-1}_i\right)\otimes\I\right]\hat{\blups}_i \\
		&= \Z_i\text{ vec}\left(\hat{\boldsymbol{\mathcal{U}}}_i\Covmat^{-1}_i\dfrac{\partial\Covmat_i}{\partial\theta_a}\right) \text{,}
	\end{split}
\end{equation}
where vec denotes column-wise vectorization and $\hat{\boldsymbol{\mathcal{U}}}_i\in\mathbb{R}^{f_i\times\eta_i}$ is the devectorized form of $\hat{\blups}_i$ so that $\text{vec}(\hat{\boldsymbol{\mathcal{U}}}_i) = \hat{\blups}_i$.
The last two lines in (\ref{eq:randomwvfast}) involve Roth's column lemma \citep{HendersonandSearle_1981, Roth_1934} which states that
\begin{equation*}
	(\mathbf{B}^T \otimes \mathbf{A})\text{vec}(\mathbf{X}) = \text{vec}(\mathbf{A}\mathbf{X}\mathbf{B}) \text{.}
\end{equation*}
Note that computing each working variate sequentially using multiple matrix-vector products as in (\ref{eq:randomwvfast}) can be relatively slow compared to a batched matrix-matrix approach:
\begin{equation*}
	\WV_{\Gparams} = \Z_i\text{ vec}\left(\hat{\boldsymbol{\mathcal{U}}}_i\Covmat^{-1}_i\D_i\right) \text{,}
\end{equation*}
where vec in this specific case reshapes $\hat{\boldsymbol{\mathcal{U}}}_i\Covmat^{-1}_i\D_i$ from an $f_i\times(\eta_ig_i)$ into an $(\eta_if_i)\times g_i$ matrix by stacking each set of $\eta_i$ columns, and $\D_i\in\mathbb{R}^{\eta_i\times(\eta_i g_i)}$ is a matrix containing all derivatives involved in random effect $i$:
\begin{equation*}
	\D_i = \begin{bmatrix}
		\dfrac{\partial\Covmat_i}{\partial\theta_1} & \dfrac{\partial\Covmat_i}{\partial\theta_a} & \dots & \dfrac{\partial\Covmat_i}{\partial\theta_{g_i}}
	\end{bmatrix} \text{.}
\end{equation*}

Working variates for parameters involved in the residual structure require residuals $\hat{\e} = \y - \Wmat_d \hat{\solution_{d}} - \Wmat_s \hat{\solution_{s}}$, $\R^{-1}$, and derivatives of $\R$:
\begin{equation*}
	\wv_b = \dfrac{\partial\R}{\partial\phi_b}\R^{-1}\hat{\e} \text{.}
\end{equation*}

\subsection{Computing the average information matrix}
Naively computing the average information matrix $\AImat = \WV^\top\Pmat\WV/2$ requires $\Pmat$.
Instead, $\AImat$ can be computed by reusing results obtained previously while solving the partitioned system (\ref{eq:MME_partitioned}).
In essence, we need to solve the multiple right-hand side system:
\begin{equation*}
	\C\mathbf{T} = \Wmat^\top \R^{-1} \WV
\end{equation*}
for $\mathbf{T} = \left[\mathbf{T}_d^\top\;\mathbf{T}_s^\top\right]^\top$.
We first solve the following system for the intermediate matrix $\boldsymbol{\Gamma}$, re-using $\LT_{ss}$:
\begin{equation*}
	\C_{ss}\boldsymbol{\Gamma} = \Wmat_s^\top \R^{-1} \WV \text{.}
\end{equation*}
We then compute the adjusted dense multiple right-hand sides:
\begin{equation*}
	\boldsymbol{\Xi} = \Wmat_d^\top \R^{-1} \WV - \C_{ds} \boldsymbol{\Gamma}
\end{equation*}
and solve the system below for the matrix of solutions corresponding to the dense equations by reusing the Cholesky factor of the Schur complement $\LT_{C/ss}$:
\begin{equation*}
	\left(\C/\C_{ss}\right)\mathbf{T}_d = \boldsymbol{\Xi}\text{.}
\end{equation*}
The matrix of solutions corresponding to the sparse equations are then computed as:
\begin{equation*}
	\mathbf{T}_s = \boldsymbol{\Gamma} - \Q \mathbf{T}_d \text{.}
\end{equation*}
Note the similarity between the process described above and the steps described in Section \ref{sec:solving}.
The main difference is that the right-hand side involving $\y$ is replaced by multiple right-hand sides involving the working variates.
The AI matrix can now be computed as:
\begin{equation*}
	\AImat = \dfrac{\WV^\top \R^{-1} \WV - \WV^\top \R^{-1} \Wmat_d \mathbf{T}_d - \WV^\top \R^{-1} \Wmat_s \mathbf{T}_s}{2}\text{,}
\end{equation*}
where the similarity with (\ref{eq:yPy}) is again easily noticed.
See Section \ref{SM:AI} of the SM for further details.

\subsection{Inversion of $\C$}
At this point the blocks of $\C^{-1}$ can be computed:
\begin{equation*}
	\C^{-1} = \begin{bmatrix}
		\C_{-dd} & \C_{-ds} \\
		\C_{-sd} & \C_{-ss}
	\end{bmatrix} \text{.}
\end{equation*}
Given the properties of the inverse of a block matrix, $\C_{-dd} = \left(\C/\C_{ss}\right)^{-1}$ is simply the inverse of the Schur complement which can be obtained from its Cholesky factor.
Then, for the off-diagonal block $\C_{-sd} = \C_{-ds}^\top = -\Q \C_{-dd}$, recalling that $\Q = \C_{ss}^{-1}\C_{sd}$.

To compute $\C_{-ss}$ we initially compute the sparse inverse subset $\C_{ss}^{-\ddagger}$ from the supernodal Cholesky factor $\LT_{ss}$ using a supernodal version of the Takahashi equations implemented in {\ttfamily cholmod-extra} \citep{Luttinen_2012}.
Note that the full inverse $\C_{ss}^{-1}$ is typically dense.
The sparse inverse subset $\C_{ss}^{-\ddagger}$ contains the subset of elements of $\C_{ss}^{-1}$ corresponding to non-zero positions in $\LT_{ss} + \LT_{ss}^\top$.
This subset is guaranteed to contain all elements of $\C_{ss}^{-1}$ that are required to compute the scores for parameters involved in random effects that are part of the sparse equations.
Using the sparse inverse subset, we compute $\C_{-ss} = \C_{ss}^{-\ddagger} + \Q \C_{-dd} \Q^\top$, noting that we only perform the addition for non-zero elements in $\C_{ss}^{-\ddagger}$.
The matrix $\C_{-ss}$ is thus a sparse inverse subset as well, with a non-zero pattern matching the one of $\LT_{ss} + \LT_{ss}^\top$.

\subsection{Computing scores}
Having obtained all elements of $\C^{-1}$ required for computing scores, we note that for each random effect $i$, given $\Covmat_{i}^{-1}\in\mathbb{R}^{\eta_i\times\eta_i}$ and $\K_{i}^{-1}\in\mathbb{R}^{f_i\times f_i}$, the associated inverse covariance matrix $\G_i^{-1} = \Covmat_{i}^{-1} \otimes \K_{i}^{-1}$ corresponds to an $\eta_i f_i\times\eta_i f_i$ diagonal block of either $\C_{-dd}$ or $\C_{-ss}$ depending on whether random effect $i$ is dense or sparse.
Denote this diagonal block $\C_{-i}$.
Next, note that this block $\C_{-i}$ can be further partitioned into $\eta_i^2$ sub-blocks of dimensions $f_i\times f_i$, such that each sub-block $\left(\C_{-i}\right)_{jj}$ corresponds to one element of $\Covmat_{i}^{-1}$.
To compute scores efficiently, we now define a matrix of traces $\mathbf{H}_i\in\mathbb{R}^{\eta_i\times\eta_i}$ for each random effect \citep{Jensenetal_1997}:
\begin{equation*}
	\mathbf{H}_i = \begin{bmatrix}
		\text{tr}\left[\K_i^{-1}\left(\C_{-i}\right)_{11}\right] & \dots & \text{tr}\left[\K_i^{-1}\left(\C_{-i}\right)_{1\eta_i}\right] \\
		\vdots & \ddots & \vdots \\
		\text{tr}\left[\K_i^{-1}\left(\C_{-i}\right)_{\eta_i1}\right] & \dots & \text{tr}\left[\K_i^{-1}\left(\C_{-i}\right)_{\eta_i\eta_i}\right]
	\end{bmatrix} \text{.}
\end{equation*}
Using $\mathbf{H}_i$, we then compute a matrix of basis-scores for random effect $i$:
\begin{equation*}
	\boldsymbol{\mathcal{S}}_i = -\dfrac{1}{2}\left(f_i\,\Covmat_i^{-1} - \Covmat_i^{-1}\mathbf{H}_i\Covmat_i^{-1} - \Covmat^{-1}_i\hat{\boldsymbol{\mathcal{U}}}_i^\top\K_i^{-1}\hat{\boldsymbol{\mathcal{U}}}_i\Covmat^{-1}_i\right) \text{,}
\end{equation*}
Finally, the random effect scores $\s_{\theta_i} = \left[s_1\dots s_a\dots s_{g_i}\right]^\top$ for each parameter $\theta_a$, associated with random effect $i$, can be computed as
\begin{equation*}
	s_a = \text{tr}\left(\dfrac{\partial\Covmat_i}{\partial\theta_a}\boldsymbol{\mathcal{S}}_i\right) \text{.}
\end{equation*}
We again use a batched approach:
\begin{equation*}
	\s_\theta = \text{vec}\left(\D_i\right)^\top \text{vec}\left(\boldsymbol{\mathcal{S}}_i\right) \text{,}
\end{equation*}
where $\text{vec}\left(\D_i\right)$ reshapes $\D_i\in\mathbb{R}^{\eta_i\times(\eta_ig_i)}$ into an $\eta_i^2\times g_i$ matrix where each column contains all elements of one of the $g_i$ derivatives.
The column-wise vectorization of $\boldsymbol{\mathcal{S}}_i$ produces a $\eta_i^2 \times 1$ vector.

Scores $\s_\phi = \left[s_1\dots s_b\dots s_r\right]^\top$ for the parameters associated with the residual structure can be obtained in similar fashion.
First, define
\begin{equation*}
	\begin{split}
		\mathbf{M}_{dd_b} &= \Wmat_d^\top \R^{-1} \dfrac{\partial\R}{\partial\phi_b} \R^{-1} \Wmat_d \\
		\mathbf{M}_{sd_b} &= \Wmat_s^\top \R^{-1} \dfrac{\partial\R}{\partial\phi_b} \R^{-1} \Wmat_d \\
		\mathbf{M}_{ss_b} &= \Wmat_s^\top \R^{-1} \dfrac{\partial\R}{\partial\phi_b} \R^{-1} \Wmat_s \text{,}
	\end{split}
\end{equation*}
after which scores can be computed as:
\begin{equation*}\label{eq:resscores}
	s_b = \dfrac{\left[\text{tr}\left(\dfrac{\partial\R}{\partial\phi_b}\R^{-1}\right) - \text{tr}\left(\mathbf{M}_{dd_b}\C_{-dd}\right) - 2\,\text{tr}\left(\mathbf{M}_{sd_b}\C_{-sd}\right) - \text{tr}\left(\mathbf{M}_{ss_b}\C_{-ss}\right) - \hat{\e}^\top \R^{-1} \dfrac{\partial\R}{\partial\phi_b} \R^{-1} \hat{\e}\right]}{-2}\text{,}
\end{equation*}
where $\C_{-ss}$ is the sparse inverse subset.
Note that the first term matches the first term of (\ref{eq:naiveresscores}), and that the three terms involving $\mathbf{M}_{dd_b}$, $\mathbf{M}_{sd_b}$, and $\mathbf{M}_{ss_b}$ together equal the second term of (\ref{eq:naiveresscores}).
See Section \ref{SM:resscores} of the SM for details on why $\hat{\e}^\top \R^{-1} \left(\partial\R/\partial\phi_b\right) \R^{-1} \hat{\e} = \y^\top\Pmat\wv_b$.

\subsection{Updating and constraining parameters}
Using the AI matrix and vector of scores
\begin{equation*}
	\scores = \left[\scores_\theta^\top\;\scores_\phi^\top\right]^\top
\end{equation*}
updates for the current parameters $\params_t$ can be computed as
\begin{equation*}
	\params_{t + 1} = \params_t + \update \text{,}
\end{equation*}
where
\begin{equation*}
	\update = \AImat^{-1} \scores \text{.}
\end{equation*}
While the quadratic nature of Newton-Raphson generally results in faster convergence compared to first-order methods such as gradient descent or expectation maximization, updates may result in parameter values that are outside the allowed space, especially when starting values are not good.
We have implemented three approaches for avoiding such updates.

First, we allow for regularization of the diagonal elements of the AI matrix corresponding to factor loadings as these are often highly correlated \citep{Gilmour_2019}:
\begin{equation*}
	\AImat^\dagger = \AImat + \AImat \circ \boldsymbol{\Pi} \text{,}
\end{equation*}
where $\boldsymbol{\Pi}$ is a diagonal matrix containing a penalty $\varrho\in[0,\infty)$ on that part of the diagonal that corresponds to loadings.

Second, we allow for damped updates such that:
\begin{equation*}
	\params_{t + 1} = \params_t + \alpha\update \text{,}
\end{equation*}
where $\alpha\in(0,1]$.

Lastly, given a proposed update $\alpha\update$, we check all parameters for excessive updates or updates that would result in parameter values outside the allowed space.
For example, variances are not allowed to in- or decrease by more than a factor $15$.
Similarly, updates for correlations or bandwidths of Gaussian kernels are not allowed to exceed an absolute value of $0.3$.
Variances and bandwidths are not allowed to become negative, and correlations must remain between $-1$ and $1$.
Loadings for a factor analytic structure cannot change by a mean magnitude of more than $10$, and any loadings that change sign cannot change by a magnitude of more than $2$.
Furthermore, loadings are estimated on the correlation scale, implying that communalities are not allowed to become $1$ or larger to ensure a positive definite correlation matrix.
Finally, any unstructured correlation matrices are checked for positive definiteness.

If any parameters are constrained given the rules above, we recompute updates for unconstrained parameters \citep{Gilmour_2019}:
\begin{equation*}
	\update_u = \AImat_{uu}^{-1} \left(\scores_u - \AImat_{uc} \update_c\right) \text{,}
\end{equation*}
where $\AImat_{uu}$ and $\AImat_{uc}$ are sub-matrices of $\AImat$ corresponding to unconstrained ($u$) and constrained ($c$) parameters.
Similarly, $\scores_u$ are scores for unconstrained parameters, and $\update_c$ are the constrained updates to be applied to constrained parameters.

Finally, we enforce a monotonic increase in the log-likelihood.
If, after potentially applying constraints, an update $\update$ results in a decrease in log-likelihood, the iteration is restarted with $\update$ halved.

While these approaches minimize convergence issues, particularly difficult scenarios may require attempting different starting values, more rigorous regularization of $\AImat$, or smaller step sizes $\alpha$, particularly for the first few iterations.

\subsection{Obtaining predictions and prediction error variances}
Given a fitted model and estimates for BLUEs and BLUPs $\hat{\solution}$, any predicted value or difference $\hat{\pi}$ may be thought of as a linear combination:
\begin{equation*}
	\hat{\pi} = \mathbf{k}^\top \hat{\solution} \text{,}
\end{equation*}
where $\mathbf{k}^\top$ is sometimes referred to as a prediction design vector \citep{Gilmour_2019, Gilmouretal_2004}.
Taking the partitioning into dense and sparse equations into account:
\begin{equation*}
	\mathbf{k} = \begin{bmatrix}
		\mathbf{k}_d \\
		\mathbf{k}_s
	\end{bmatrix}
\end{equation*}
and
\begin{equation*}
	\hat{\solution} = \begin{bmatrix}
		\hat{\solution}_d \\
		\hat{\solution}_s
	\end{bmatrix} \text{.}
\end{equation*}
Note that
\begin{equation*}
	\begin{split}
		\text{Var}\left(\hat{\pi}\right) &= \mathbf{k}^\top \text{Var}\left(\hat{\solution}\right) \mathbf{k} \\
		&= \mathbf{k}^\top \C^{-1} \mathbf{k} \text{,}
	\end{split}
\end{equation*}
which can be expanded as follows:
\begin{equation*}
	\begin{split}
		\text{Var}\left(\hat{\pi}\right) &= \begin{bmatrix}
			\mathbf{k}_d^\top & \mathbf{k}_s^\top
		\end{bmatrix} \begin{bmatrix}
			\C_{-dd} & \C_{-ds} \\
			\C_{-sd} & \C_{-ss}
		\end{bmatrix} \begin{bmatrix}
			\mathbf{k}_d \\
			\mathbf{k}_s
		\end{bmatrix} \\
		&= \begin{bmatrix}
			\mathbf{k}_d^\top & \mathbf{k}_s^\top
		\end{bmatrix} \begin{bmatrix}
			\left(\C/\C_{ss}\right)^{-1} & -\left(\C/\C_{ss}\right)^{-1}\Q^\top \\
			-\Q \left(\C/\C_{ss}\right)^{-1} & \C_{ss}^{-1} + \Q \left(\C/\C_{ss}\right)^{-1} \Q^\top
		\end{bmatrix} \begin{bmatrix}
			\mathbf{k}_d \\
			\mathbf{k}_s
		\end{bmatrix} \\
		&= \mathbf{k}_d^\top \left(\C/\C_{ss}\right)^{-1} \mathbf{k}_d - \mathbf{k}_s^\top \Q \left(\C/\C_{ss}\right)^{-1} \mathbf{k}_d \\
		&\quad - \mathbf{k}_d^\top \left(\C/\C_{ss}\right)^{-1} \Q^\top \mathbf{k}_s + \mathbf{k}_s^\top \C_{ss}^{-1} \mathbf{k}_s + \mathbf{k}_s^\top \Q \left(\C/\C_{ss}\right)^{-1} \Q^\top \mathbf{k}_s \\
		&= \mathbf{k}_s^\top \C_{ss}^{-1} \mathbf{k}_s + \left(\mathbf{k}_d - \Q^\top \mathbf{k}_s\right)^\top \left(\C/\C_{ss}\right)^{-1} \left(\mathbf{k}_d - \Q^\top \mathbf{k}_s\right) \text{.}
	\end{split}
\end{equation*}
The first term can be easily obtained by solving the triangular system:
\begin{equation*}
	\LT_{ss}\boldsymbol{\gamma} = \mathbf{k}_s
\end{equation*}
and then computing:
\begin{equation*}
	\mathbf{k}_s^\top \C_{ss}^{-1} \mathbf{k}_s = \boldsymbol{\gamma}^\top \boldsymbol{\gamma} \text{.}
\end{equation*}
The second term involves $\Q$ and $\left(\C/\C_{ss}\right)^{-1} = \C_{-dd}$ which are both readily available.

\section{Extensions}\label{sec:extensions}
The algorithm outlined in Section \ref{sec:algo} lends itself well to some extensions related to solving both the dense and sparse equations, as well as overall model specification.
These extensions are detailed below.

\subsection{GPU accelerated solving of dense equations}
Given a fully dense model or set of dense equations, the use of a parallel computing framework like NVIDIA's CUDA can accelerate solving the system significantly.
Our implementation involves a number of steps, divided into four phases.
In terms of notation, we will describe the case of a mix of dense and sparse equations.
The first phase involves initializing CUDA by allocating memory on the GPU for the dense Schur complement $\C/\C_{ss}$, $\solution_{d}$, $\mathbf{T}_d$, $\Q^\top$, as well as a vector for the diagonal of $\LT_{ss}$ and the dense matrix $\Q\C_{-dd}$.
We also allocate memory for a {\ttfamily cuSOLVER} workspace buffer equal to the maximum of the values computed using {\ttfamily cusolverDnDpotrf\_bufferSize()} and {\ttfamily cusolverDnDpotri\_bufferSize()}, as well as a {\ttfamily cuSPARSE} buffer computed using {\ttfamily cusparseSpMM\_bufferSize}.
Then, for the second phase:
\begin{enumerate}
	\item Copy $\C/\C_{ss}$ and $\boldsymbol{\xi}$ to the GPU.
	\item Compute $\LT_{C/ss}$ in-place using {\ttfamily cusolverDnDpotrf()}, overwriting $\C/\C_{ss}$.
	\item Solve $\C/\C_{ss}\solution_{d}=\boldsymbol{\xi}$ for $\solution_{d}$ in-place using {\ttfamily cusolverDnDpotrs()}, overwriting $\boldsymbol{\xi}$.
	\item Copy $\solution_{d}$ and the diagonal elements of $\LT_{C/ss}$ back to the CPU.
	\item Compute $\text{log}|\C/\C_{ss}|$ using the diagonal elements of $\LT_{C/ss}$.
\end{enumerate}
At this point, the algorithm continues on the CPU to compute, among other things, $\solution_{s}$, the log-likelihood, and working variates.
Crucially, $\LT_{C/ss}$ is kept alive in GPU memory to be reused for solving against the working variates in the third phase:
\begin{enumerate}
	\item Copy $\Q^\top$ and $\boldsymbol{\Xi}$ to the GPU.
	\item Solve $\C/\C_{ss}\mathbf{T}_{d}=\boldsymbol{\Xi}$ for $\mathbf{T}_{d}$ in-place using {\ttfamily cusolverDnDpotrs()}, overwriting $\boldsymbol{\Xi}$.
	\item Compute the inverse $\left(\C/\C_{ss}\right)^{-1}$ in-place from the Cholesky factor using {\ttfamily cusolverDnDpotri()}, overwriting $\LT_{C/ss}$.
	\item Compute the sparse-dense matrix product $\Q\left(\C/\C_{ss}\right)^{-1}$ using {\ttfamily cusparseSpMM()}. Note that we copy $\Q^\top$ to the GPU rather than $\Q$ as $\Q$ is stored in compressed sparse column (CSC) format while {\ttfamily cusparseSpMM()} is typically faster when used with a sparse matrix stored in compressed sparse row (CSR) format. The CSC memory arrays of $\Q^\top$ are identical to the CSR format of $\Q$ after swapping dimensions.
	\item Copy $\mathbf{T}_{d}$, $\left(\C/\C_{ss}\right)^{-1}$, and $\Q\left(\C/\C_{ss}\right)^{-1}$ back to the CPU.
\end{enumerate}
The final phase involves destroying all objects left on the GPU and freeing memory after the algorithm has converged or stopped for some other reason.
This approach delegates most work involved in solving the dense equations to the GPU, significantly speeding up model fitting.

\subsection{Decoupling estimation of correlation and variance structures}
Recall that the covariance matrix for any random effect can be denoted as:
\begin{equation*}
	\G = \Covmat \otimes \K
\end{equation*}
and that $\Covmat$ itself may be a Kronecker product of several structures, e.g., for modeling three factors in the data, one of which involves a known structure, so that with $\Covmat = \Covmat_1 \otimes \Covmat_2$:
\begin{equation*}
	\G = \Covmat_1 \otimes \Covmat_2 \otimes \K_3 \text{,}
\end{equation*}
where $\Covmat_1\in\mathbb{R}^{p\times p}$ and $\Covmat_2\in\mathbb{R}^{q\times q}$.
Current implementations are limited in the choices of how variances are modeled.
By default, the software package {\ttfamily ASReml-R}, for example, imposes the restriction that only one structure may be estimated on the covariance scale, resulting in either $p$ or $q$ variances \citep{butleretal_2023}.
Using {\ttfamily ASReml-R}'s {\ttfamily own()} function allows specification of less restrictive variance models, but requires manually defining all derivative matrices and is restricted to dense models \citep{Melsenetal_2026}.
The {\ttfamily R}-package {\ttfamily sommer} slightly relaxes this restriction by allowing for a strictly multiplicative variance structure through the estimation of an overall variance, $p-1$ variance ratios for $\Covmat_1$, and $q-1$ variance ratios for $\Covmat_2$, resulting in $p + q - 1$ variance parameters \citep{Covarrubias-Pazaran_2016}.
Our implementation completely decouples estimation of correlation matrices and variances, allowing for a full interaction structure on the latter:
\begin{equation*}
	\G = \boldsymbol{\sigma}\boldsymbol{\sigma}^\top \circ \left(\Covmat_1 \otimes \Covmat_2\right) \otimes \K_3 \text{,}
\end{equation*}
where $\boldsymbol{\sigma}\in\mathbb{R}^{pq}$ is a column vector containing unique standard deviations for all combinations of levels of factors $1$ and $2$.
Note that $\Covmat_1$ and $\Covmat_2$ are both estimated on the correlation scale.

\subsection{Stacking conformable covariance structures}
Denote the covariance matrices of two random effects involving the same known structure as:
\begin{equation*}
	\begin{split}
		\G_1 &= \Covmat_1 \otimes \K \\
		\G_2 &= \Covmat_2 \otimes \K \text{.}
	\end{split} 
\end{equation*}
If both $\Covmat_1$ and $\Covmat_2$ are used to model covariances among levels of the same factor, e.g., to model G$\times$E interactions using both a Gaussian kernel and factor analytic structure as proposed by \citet{Melsenetal_2026}, they can easily be stacked in a single random effect:
\begin{equation*}
		\G_{1+2} = \left(\Covmat_1 + \Covmat_2\right) \otimes \K \text{.}
\end{equation*}
Note that this idea was also briefly mentioned by \citet{Gilmour_2019}, and only works if design matrices $\Z_1$ and $\Z_2$ are equivalent as when $\Z_1 = \Z_2$:
\begin{equation*}
	\Z_1\left(\Covmat_1 \otimes \K\right)\Z_1^\top + \Z_2\left(\Covmat_2 \otimes \K\right)\Z_2^\top + \R = \Z\left[\left(\Covmat_1 + \Covmat_2\right)\otimes\K\right]\Z^\top + \R \text{.}
\end{equation*}

\section{Illustrations}\label{sec:illustrations}
The following sections aim to illustrate the use of our hybrid AI REML implementation, as well as the extensions.
We start by fitting a few basic multivariate linear mixed models.
We then show the benefit of GPU acceleration using a large GBLUP model, followed by an application involving the stacking of covariance matrices, and finally a single-stage kernel-based G$\times$E$\times$M model involving both dense and sparse equations.
Comparisons were made with {\ttfamily kmmr} version {\ttfamily ea446b9}, {\ttfamily ASReml-R} version {\ttfamily 4.2}, and {\ttfamily sommer} version {\ttfamily 4.4.88} \citep{butleretal_2023, Covarrubias-Pazaran_2016}.
All models were fitted on a machine with $64$GB of RAM and an Intel i7-1370P CPU with Intel oneMKL BLAS and LAPACK version {\ttfamily 2025.2.0} \citep{Intel_2025}.
GPU accelerated models were fitted on a node with $64$GB of RAM, NVIDIA $80$GB A100 GPU, and 8 Intel Skylake CPU cores.

\subsection{Basic G$\times$E models}
We use the DROPS (Drought-tolerant yielding plants, \citealt{milletetal_2019, milletetal_2016}) dataset to illustrate the fitting of some basic models.
The dataset contains BLUEs with weights for $234$ maize hybrids, $14$ environments (location-years), $2$ managements, as well as plot-level data and a SNP-based kinship matrix.

\subsubsection{Factor analytic models for G$\times$E with weights}
The first model we fit is the standard extended (sparse) factor analytic model \citep{Thompsonetal_2003}.
Given the mixed model in (\ref{eq:LMM}):
\begin{equation}\label{eq:drops:fa}
	\mathbf{u} \sim \mathcal{N}\left(\mathbf{0}, \Covmat \otimes \K\right) \text{,}
\end{equation}
where $\K$ is the kinship matrix and:
\begin{equation}\label{eq:exfa}
	\Covmat = \begin{bmatrix}
		\LA\LA^\top + \PS & -\LA \\
		-\LA^\top & \I
	\end{bmatrix}
\end{equation}
models the $28$ combinations of managements and environments, as well as the $m$ latent factors.
The design matrix $\X$ contains main effects for all $28$ E$\times$M combinations.
Residuals are assumed independent and the residual variance for each observation is fixed according to the corresponding weight from the first stage analyses using the diagonal approximation \citep{Damesaetal_2017}.

Table \ref{tbl:fa-drops} shows some results for both $1$ and $3$ factors.
Log-likelihoods are practically equal between {\ttfamily ASReml-R} and {\ttfamily kmmr}, with {\ttfamily sommer} reporting a value deviating by some constant.
Runtimes per iteration are slowest for {\ttfamily ASReml-R}, followed by {\ttfamily sommer} and then {\ttfamily kmmr}.
Total runtime for {\ttfamily sommer} was slightly longer than for {\ttfamily ASReml-R} in the FA1 case due to the larger number of iterations required to reach convergence.
Multi-threading led to a small decrease in runtimes for both {\ttfamily sommer} and the FA3 {\ttfamily kmmr} model, although speedups for the latter were negligible.
For the FA1 model, multi-threading actually led to a slight increase in runtime for {\ttfamily kmmr}.
BLUPs from all models agreed, with correlations between BLUPs consistently larger than $0.999$.
\begin{table}
	\caption{Runtime in seconds and log-likelihood values for order $1$ and $3$ sparse factor analytic models. The number of CPU threads is denoted by $t$. By default, {\ttfamily ASReml-R} uses as many threads as there are available. Fastest times are shown in bold.}
	\label{tbl:fa-drops}
	\begin{tabular}{ cccccc }
		Software & Settings & Runtime & Iterations & Time per iteration & Log-likelihood \\
		\hline
		ASReml-R & FA1 & 17.98 & 13 & 1.38 & -2038.877 \\
		sommer, $t$=1 & FA1 & 26.31 & 22 & 1.20 & 499.777 \\
		kmmr, $t$=1 & FA1 & \textbf{4.95} & 16 & \textbf{0.31} & -2038.876 \\
		sommer, $t$=8 & FA1 & 22.92 & 22 & 1.04 & 499.777 \\
		kmmr, $t$=8 & FA1 & 5.58 & 16 & 0.35 & -2038.876 \\
		\hline
		ASReml-R & FA3 & 149.39 & 39 & 3.83 & -1637.879 \\
		sommer, $t$=1 & FA3 & 77.78 & 53 & 1.47 & 900.768 \\
		kmmr, $t$=1 & FA3 & 22.12 & 23 & 0.96 & -1637.890 \\
		sommer, $t$=8 & FA3 & 69.57 & 53 & 1.31 & 900.768 \\
		kmmr, $t$=8 & FA3 & \textbf{21.08} & 23 & \textbf{0.92} & -1637.890 \\
	\end{tabular}
\end{table}

\subsubsection{Unstructured model and Gaussian kernel for G$\times$E$\times$M with weights}
The second model we fitted was the MVGK model introduced by \citet{Melsenetal_2026}.
Models again contained fixed main effects for all management by environment combinations.
Furthermore,
\begin{equation}\label{eq:MVGK}
	\mathbf{u} \sim \mathcal{N}\left[\mathbf{0}, \mathbf{ss}^\top\circ\left(\Covmat \otimes e^{-h\D}\right) \otimes \K\right] \text{,}
\end{equation}
where $\Covmat$ is an unstructured correlation matrix for managements, $\D$ is a squared Euclidean distance matrix of environments, $h$ is a strictly positive bandwidth, and $\mathbf{s}$ contains standard deviations for all E$\times$M combinations.
Weights were used as before for residual variances.

{\ttfamily ASReml-R} was again slowest both in total runtime and time per iteration, while log-likelihoods were similar (Table \ref{tbl:usgk-drops}).
The benefits of multi-threading and GPU acceleration are clear, given that the fitted G$\times$E$\times$M covariance structure has a dense inverse.
We do want to stress the importance of appropriate starting values as they heavily impacted the number of iterations required for {\ttfamily kmmr} to reach convergence.
Given inappropriate starting values, convergence slowed down dramatically close to the maximum likelihood, bringing total runtime close to {\ttfamily ASReml-R}.
We do note that this is not a computational issue, but rather a statistical issue that requires further investigation.
Correlations between BLUPs from the different models were again larger than $0.999$.
\begin{table}
	\caption{Runtime in seconds and log-likelihood values for combinations of unstructured and Gaussian kernel models for managements and environments, respectively. The number of CPU threads is denoted by $t$. By default, {\ttfamily ASReml-R} uses as many threads as there are available. Fastest times are shown in bold.}
	\label{tbl:usgk-drops}
	\begin{tabular}{ cccccc }
		Software & Settings & Runtime & Iterations & Time per iteration & Log-likelihood \\
		\hline
		ASReml-R & MVGK & 365.92 & 12 & 30.49 & -1915.114 \\
		kmmr, $t$=1 & MVGK & 76.23 & 13 & 5.86 & -1915.119 \\
		kmmr, $t$=8 & MVGK & 40.73 & 13 & 3.13 & -1915.119 \\
		kmmr, GPU & MVGK & \textbf{12.29} & 13 & \textbf{0.95} & -1915.119 \\
	\end{tabular}
\end{table}

\subsubsection{Single-stage models for G$\times$E}
The third illustration involves a single-stage model for the $14$ rain-fed trials in the DROPS dataset.
These models contain fixed main effects for all environments, as well as i.i.d. environment-specific random effects for complete blocks and incomplete blocks within complete blocks.
They also contain environment-specific random effects with a separable order $1$ autoregressive (AR1) structure for rows and columns, with a single variance component.
Residuals were assumed to be independent with environment-specific nugget residual variances.
Finally, the random nested G$\times$E effect is as described in (\ref{eq:drops:fa}).

Table \ref{tbl:ss-drops} shows that while {\ttfamily kmmr} is faster in terms of both total runtime and time per iteration, it fails to converge to the exact same log-likelihood maximum as {\ttfamily ASReml-R} for both the FA1 and FA3 models.
While correlations between G$\times$E BLUPs from both models were again consistently higher than $0.999$, it indicates that our strategies for choosing starting values or parameter constraining may be sub-optimal.
Interestingly, multi-threading actually slowed down {\ttfamily kmmr}, possibly due to the highly sparse structure of the coefficient matrix and lack of large supernodes where multi-threaded BLAS and LAPACK would bring a benefit.
\begin{table}
	\caption{Runtime in seconds and log-likelihood values for order $1$ and $3$ single stage models involving a sparse factor analytic nested G$\times$E effect. The number of CPU threads is denoted by $t$. By default, {\ttfamily ASReml-R} uses as many threads as there are available. Fastest times are shown in bold.}
	\label{tbl:ss-drops}
	\begin{tabular}{ cccccc }
		Software & Settings & Runtime & Iterations & Time per iteration & Log-likelihood \\
		\hline
		ASReml-R & FA1 & 203.08 & 34 & 5.97 & -3937.646 \\
		kmmr, $t$=1 & FA1 & \textbf{34.84} & 25 & \textbf{1.39} & -3948.664 \\
		kmmr, $t$=8 & FA1 & 39.90 & 25 & 1.60 & -3948.664 \\
		\hline
		ASReml-R & FA3 & 342.06 & 47 & 7.28 & -3763.711 \\
		kmmr, $t$=1 & FA3 & \textbf{80.33} & 42 & \textbf{1.91} & -3773.896 \\
		kmmr, $t$=8 & FA3 & 90.21 & 42 & 2.15 & -3773.896 \\
	\end{tabular}
\end{table}

\subsection{The benefit of GPU acceleration}
We illustrate the benefit of GPU acceleration using a large single-trial Australian wheat dataset containing spatially corrected phenotypes for $10375$ genotypes as well as the DROPS dataset described previously.

\subsubsection{Large GBLUP models}
The benefit of using GPU acceleration is most obvious when dealing with large, purely dense systems.
This illustration involves fitting a simple GBLUP model to the 2014 Roseworth wheat trial data \citep{Yanetal_2023, Normanetal_2017}:
\begin{equation*}
	\mathbf{u} \sim \mathcal{N}\left(\mathbf{0}, \sigma^2 \K\right) \text{,}
\end{equation*}
where $\K\in\mathbb{R}^{10375\times 10375}$ is the GRM with dense inverse and $\sigma^2$ is the additive genetic variance.
Residuals were assumed to be i.i.d.

Table \ref{tbl:AGT} shows the significant time savings obtained by fitting these dense models on a GPU.
Note that at this point, computing the log determinant of $\K$ in the first iteration takes up a significant proportion of the total runtime, especially for the GPU accelerated solver.
Comparisons excluding the time required for $\text{log}|\K|$ may be more useful as this expensive log determinant is computed once and reused throughout the remaining iterations.
\begin{table}
	\caption{Runtime in seconds and log-likelihood values for the $10375$ genotype GBLUP model. The number of CPU threads is denoted by $t$. By default, {\ttfamily ASReml-R} uses as many threads as there are available.
		Approximate runtimes excluding the one-time calculation of the log-determinant of $\K$ are shown between parentheses. Fastest times are shown in bold.}
	\label{tbl:AGT}
	\begin{tabular}{ ccccc }
		Software & Runtime & Iterations & Time per iteration & Log-likelihood \\
		\hline
		ASReml-R & 988.60 (878.72) & 8 & 123.58 (109.84) & -69106.05 \\
		sommer, $t$=1 & 511.49 (342.00) & 9 & 56.83 (38.00) & -3966.746 \\
		kmmr, $t$=1 & 301.81 (223.71) & 11 & 27.44 (20.34) & -69106.054 \\
		sommer, $t$=8 & 362.17 (169.00) & 9 & 40.24 (21.13) & -3966.746 \\
		kmmr, $t$=8 & 129.33 (109.81) & 11 & 11.76 (9.98) & -69106.054 \\
		kmmr, GPU & \textbf{49.16} (\textbf{14.86}) & 11 & \textbf{4.47} (\textbf{1.35}) & -69106.054 \\
	\end{tabular}
\end{table}

\subsubsection{Stacking covariance matrices}
The next illustration involves fitting both a multiple-variance Gaussian kernel as well as an order $2$ factor analytic structure to the rain-fed trials from the DROPS dataset.
Weights were again used for residual variances and all models contained fixed environment main effects.
The random effects can be fitted separately:
\begin{equation*}
	\begin{split}
		\mathbf{u}_{FA} &\sim \mathcal{N}\left(\mathbf{0}, \Covmat_{FA}\otimes\K\right) \\
		\mathbf{u}_{GK} &\sim \mathcal{N}\left(\mathbf{0}, \Covmat_{GK}\otimes\K\right) \text{,}
	\end{split}
\end{equation*}
where $\Covmat_{FA}$ has the structure described in (\ref{eq:exfa}) and $\Covmat_{GK} = \mathbf{ss}^\top \circ e^{-h\D}$.
Alternatively, both effects can be stacked resulting in a single random effect:
\begin{equation*}
	\mathbf{u}_{FA + GK} \sim \mathcal{N}\left[\mathbf{0}, \left(\Covmat_{FA} + \Covmat_{GK}\right)\otimes\K\right] \text{,}
\end{equation*}
where:
\begin{equation*}
	\begin{split}
		\Covmat_{FA} &= \mathbf{s}_{FA}\mathbf{s}_{FA}^\top \circ \left(\LA\LA^\top + \PS\right) \\
		\Covmat_{GK} &= \mathbf{s}_{GK}\mathbf{s}_{GK}^\top \circ e^{-h\D} \text{,}
	\end{split}
\end{equation*}
where loadings $\LA$ are estimated on the correlation scale, diagonal matrix with unique variances $\PS$ follows from the diagonal elements of $\LA\LA^\top$, $\mathbf{s}_{FA}$ contains standard deviations associated with the factor analytic structure, and $\mathbf{s}_{GK}$ contains standard deviations associated with the Gaussian kernel.

The benefit of stacking conformable covariance structures is especially clear (Table \ref{tbl:stack_drops}).
Times per iteration were reduced significantly, even for the single-threaded {\ttfamily kmmr} model.
Multi-threading and GPU acceleration further accelerated model fitting.
Note that {\ttfamily ASReml-R} failed to converge after $50$ iterations, moving around the log-likelihood value of $-630.802$.
This is likely again an issue with parameter constraining or choosing starting values, and not of practical significance as the correlation between $\mathbf{u}_{FA + GK}$ from {\ttfamily kmmr} and $\mathbf{u}_{FA} + \mathbf{u}_{GK}$ from {\ttfamily ASReml-R} was again larger than $0.999$.

\begin{table}
	\caption{Runtime in seconds and log-likelihood values for models involving both (sparse) factor analytic and Gaussian kernel-based nested G$\times$E effects. The number of CPU threads is denoted by $t$. By default, {\ttfamily ASReml-R} uses as many threads as there are available.
	Note that the model fitted using {\ttfamily ASReml-R} contained two random effects, while the models implemented in {\ttfamily kmmr} contain a single stacked effect.
	$^*$Note that {\ttfamily ASReml-R} failed to converge within $50$ iterations. Fastest times are shown in bold.}
	\label{tbl:stack_drops}
	\begin{tabular}{ ccccc }
		Software & Runtime & Iterations & Time per iteration & Log-likelihood \\
		\hline
		ASReml-R (2 random effects) & 1036.48 & 50 & 20.73 & -630.802$^*$ \\
		kmmr, stacked, $t$=1 & 18.48 & 20 & 0.92 & -616.340 \\
		kmmr, stacked, $t$=8 & 12.07 & 20 & 0.60 & -616.340 \\
		kmmr, stacked, GPU & \textbf{6.79} & 20 & \textbf{0.34} & -616.340 \\
	\end{tabular}
\end{table}

\subsubsection{Single stage G$\times$E$\times$M using Gaussian kernel}
The final illustration shows that GPU acceleration also brings benefits for models containing both dense equations as well as a significant number of sparse equations.
Models contained fixed main effects for all combinations of managements and environments.
We fitted i.i.d. random effects for complete blocks and incomplete blocks within complete blocks for every E$\times$M combination.
Similarly, models contained random effects with a separable AR1$\times$AR1 structures for rows and columns at each E$\times$M combination.
The residual covariance structure was assumed to be diagonal with heterogeneous variances.
The nested random G$\times$E$\times$M effect was modeled using the covariance matrix as defined in (\ref{eq:MVGK}).

Table \ref{tbl:ssGEM_drops} shows that GPU acceleration remains beneficial even for models containing both dense and sparse equations, providing and approximately two-fold speedup.
Note that while changes in the log-likelihood became small, the models did not reach convergence after 50 iterations.
Finally, we want to point out that the rather slow Intel Skylake CPU on the A100 node we used for the GPU accelerated model may have had an impact here.
We reran the single-threaded CPU solver as well as the GPU accelerated solver on a node equipped with an NVIDIA L40S GPU and 8 AMD Zen 5 Turin CPU cores (64GB RAM) and achieved runtimes of $17.85$ and $7.60$ seconds per iteration, respectively.
In any case, the benefit of our hybrid GPU/CPU implementation is clear.
\begin{table}
	\caption{Runtime in seconds and log-likelihood values for single stage models involving an unstructured correlation matrix for managements and Gaussian kernel for environments. The number of CPU threads is denoted by $t$. $^*$No convergence after 50 iterations. Fastest times are shown in bold.}
	\label{tbl:ssGEM_drops}
	\begin{tabular}{ ccccc }
		Software & Runtime & Iterations & Time per iteration & Log-likelihood \\
		\hline
		kmmr, $t$=1 & 1431.01 & 50 & 28.62 & -7337.195$^*$ \\
		kmmr, $t$=8 & 1286.48 & 50 & 25.73 & -7337.195$^*$ \\
		kmmr, GPU & \textbf{675.88} & 50 & \textbf{13.52} & -7337.195$^*$ \\
	\end{tabular}
\end{table}

\section{Code and data availability}
All data and code required to reproduce the presented results is available at \url{https://github.com/KillianMelsen/kmmr-benchmarking}, with the exception of the kinship and SNP marker data for the 2014 Roseworth trial due to file size limits.
Raw data, including SNP markers, can be found in the original publications regarding DROPS \citep{milletetal_2016, milletetal_2019} and the 2014 Roseworth trial \citep{Yanetal_2023, Normanetal_2017}.
Source code for our hybrid GPU/CPU AI REML implementation is available at \url{https://github.com/KillianMelsen/kmmr}.

\section{Discussion}\label{sec:discussion}
We have shown how GPU acceleration and supernodal Cholesky factorization can significantly accelerate the fitting of complex linear mixed models.
The benefits of using a GPU are most clear when coefficient matrices contain large dense blocks as a result of SNP-based GRMs or kernels constructed from environmental features.
On the other hand, our software also allows for modeling random effects as part of the sparse equations when the structure allows.
Common examples of such structures include effects with a diagonal covariance structure such as blocks or other trial design factors, as well as AR1 structures.

The importance of facilitating both dense and sparse structures is likely to increase as the large scale collection of genomic and environmental data continues \citep{Crossaetal_2021}.
As mentioned, unlike pedigree-based relationship matrices, the inverses of GRMs are dense which significantly increases computational burden \citep{Meyer_2023}.
Similarly, kernels constructed from environmental features typically have dense inverses \citep{Melsenetal_2026}.
While linear kernels may potentially be approximated using an extended factor analytic structure of which the parameters are considered fixed, no such straightforward approximation is available for the Gaussian kernels.

There are limits to applying GPU acceleration to the dense equations, however.
Memory requirements of storing the dense Schur complement scale quadratically with the number of dense equations, and in addition to the Schur complement itself, memory must be allocated for temporary work buffers.
For large datasets with many genotypes and environments, the number of equations, dimensions of the coefficient matrix, and thus storage requirements may simply be too large.

Often, however, the datasets involving a large number of equations are sparse in the sense that not all genotypes have been phenotyped in every environment \citep{G2F_2025, Jarquinetal_2020}.
In these cases, \citet{Melsenetal_2026} showed that the use of Gaussian kernels is especially useful for modeling G$\times$E$\times$M interactions.
As for such datasets the number of equations can be much larger than the number of phenotypic records, the direct inversion formulation of the AI REML algorithm may be a better alternative.
Rather than forming the large coefficient matrix and solving the MMEs, the direct inversion formulation works in the observation space by factorizing $\V = \Z\G\Z^\top + \R$ instead of $\C$ \citep{LeeandvanderWerf_2006}.
This direct inversion algorithm has been implemented in different software packages for genomic analyses, including {\ttfamily MTG2} and {\ttfamily sommer} \citep{Covarrubias-Pazaran_2016, LeeandvanderWerf_2016}.
A GPU accelerated implementation of this algorithm would be a natural extension of the work presented here.
As $\V$ is generally dense, its factorization, any required solves, or multiplications are well suited to the use of a GPU.

There are also opportunities with regards to improving the sparse implementation.
We currently relied on {\ttfamily CHOLMOD} for the supernodal Cholesky factorization and {\ttfamily cholmod-extra} for computing the sparse inverse subset \cite{Chenetal_2008, Luttinen_2012}.
\cite{Mazur_2023} compared several options for computing the supernodal Cholesky factor and subsequently the sparse inverse subset.
The comparisons suggest that the choice of sparse solver and inversion algorithm can have a significant impact on the total amount of time required.
Our reliance on {\ttfamily CHOLMOD} and {\ttfamily cholmod-extra} is by no means a requirement, and it may be worthwhile to explore other packages.
Of these alternatives, {\ttfamily MUMPS} (multifrontal massively parallel sparse direct solver) may be an interesting option \citep{Amestoyetal_2001}.
It relies on a multifrontal instead of left-looking supernodal Cholesky factorization and, as shown by \citet{Mazur_2023}, may be a better choice for the coefficient matrices often encountered in the analysis of plant breeding data.

Regardless of the specific solvers used, the results we present here show that within linear mixed models, exploiting sparsity and high-throughput parallel dense computing strategies are not exclusive, but rather complimentary.
By relying on the optimal approach for the different sets of equations, we consistently achieve computing times that are lower than alternative implementations, while retaining the flexibility of AI REML.






\newpage
\begin{center}
	\begin{minipage}{\linewidth}
		\centering
		\textbf{\large Supplementary Material: A hybrid GPU/CPU implementation of the average information restricted maximum likelihood algorithm for fitting multivariate linear mixed models}
		\vskip0.5cm
		Killian A.C.\ Melsen$^{1*}$, Fred A.\ van Eeuwijk$^1$, Carel F.W.\ Peeters$^1$
	\end{minipage}
	\vskip0.5cm
	\begin{minipage}{\linewidth}
		\raggedright
		$^1$Mathematical \& Statistical Methods group - Biometris, Wageningen University \& Research, PO Box 16, 6700 AA, Wageningen, The Netherlands\\
		$^*$ \url{killian.melsen@wur.nl}
	\end{minipage}
	\vskip2cm
\end{center}
\setcounter{equation}{0}
\setcounter{figure}{0}
\setcounter{table}{0}
\setcounter{page}{1}
\setcounter{section}{0}
\makeatletter
\renewcommand{\theequation}{S\arabic{equation}}
\renewcommand{\thefigure}{S\arabic{figure}}
\renewcommand{\bibnumfmt}[1]{[S#1]}
\renewcommand{\citenumfont}[1]{S#1}
\renewcommand{\thetable}{S\arabic{table}}
\renewcommand{\thesection}{S.\arabic{section}}
\section{Solving the partitioned system}\label{SM:solving}
Section \ref{sec:solving} of the main text deals with solving the following partitioned system:
\begin{equation*}
	\begin{bmatrix}
		\C_{dd} & \C_{ds} \\
		\C_{sd} & \C_{ss}
	\end{bmatrix}\begin{bmatrix}
		\solution_d \\
		\solution_s
	\end{bmatrix} = \begin{bmatrix}
		\Wmat_d^\top \R^{-1} \y \\
		\Wmat_s^\top \R^{-1} \y
	\end{bmatrix} \text{.}
\end{equation*}
Note that the two sets of equations can be written as
\begin{equation}\label{SM:solving:1}
	\begin{split}
		&\C_{dd} \solution_d + \C_{ds} \solution_s = \Wmat_d^\top \R^{-1} \y \\
		&\C_{sd} \solution_d + \C_{ss} \solution_s = \Wmat_s^\top \R^{-1} \y
	\end{split}
\end{equation}
and that thus, by rewriting the second set of equations:
\begin{equation*}
	\C_{ss} \solution_s = \Wmat_s^\top \R^{-1} \y - \C_{sd} \solution_d
\end{equation*}
giving
\begin{equation}\label{SM:solving:2}
	\begin{split}
		\solution_s &= \C_{ss}^{-1} \left(\Wmat_s^\top \R^{-1} \y - \C_{sd} \solution_d\right) \\
		&= \C_{ss}^{-1} \Wmat_s^\top \R^{-1} \y - \C_{ss}^{-1} \C_{sd} \solution_d \text{.}
	\end{split}
\end{equation}
Recall that we defined $\Q = \C_{ss}^{-1}\C_{sd}$.
Substituting the definition of $\solution_s$ from (\ref{SM:solving:2}) into the first set of equations of (\ref{SM:solving:1}) gives:
\begin{equation*}
	\C_{dd} \solution_d + \C_{ds} \C_{ss}^{-1} \Wmat_s^\top \R^{-1} \y - \C_{ds} \Q \solution_d = \Wmat_d^\top \R^{-1} \y \text{,}
\end{equation*}
which can be rearranged to produce the system:
\begin{equation}\label{SM:solving:3}
	\left(\C_{dd} - \C_{ds}\Q\right)\solution_d = \Wmat_d^\top\R^{-1}\y - \C_{ds}\C_{ss}^{-1}\Wmat_s^\top\R^{-1}\y \text{,}
\end{equation}
where $\C_{dd} - \C_{ds}\Q = \C/\C_{ss}$ is the Schur complement of $\C_{ss}$ in $\C$.
Now solve the intermediate system
\begin{equation*}
	\C_{ss}\boldsymbol{\gamma} = \Wmat_s^\top\R^{-1}\y
\end{equation*}
for $\boldsymbol{\gamma}$.
Then, the solutions to the dense equations can be obtained by solving system (\ref{SM:solving:3}), substituting $\boldsymbol{\gamma}$:
\begin{equation*}
	\solution_d = \left(\C/\C_{ss}\right)^{-1}\left(\Wmat_d^\top\R^{-1}\y - \C_{ds}\boldsymbol{\gamma}\right) \text{.}
\end{equation*}
The dense solutions $\solution_d$ can then be used to easily obtain solutions to the sparse equations using (\ref{SM:solving:2}), $\Q$, and $\boldsymbol{\gamma}$:
\begin{equation*}
	\solution_s = \boldsymbol{\gamma} - \Q \solution_d \text{.}
\end{equation*}

\section{The projection matrix}\label{SM:P}
To see why
\begin{equation}\label{SMAIstart}
	\R^{-1} - \R^{-1}\Wmat\C^{-1}\Wmat^\top\R^{-1} = \Pmat \text{,}
\end{equation}
define
\begin{equation*}
	\C^{-1} = \begin{bmatrix}
		\X^\top \R^{-1} \X & \X^\top \R^{-1} \Z \\
		\Z^\top \R^{-1} \X & \Z^\top \R^{-1} \Z + \G^{-1}
	\end{bmatrix}^{-1} = \begin{bmatrix}
		\C_{-xx} & \C_{-xz} \\
		\C_{-zx} & \C_{-zz}
	\end{bmatrix} \text{,}
\end{equation*}
and $\C_{zz} = \Z^\top \R^{-1} \Z + \G^{-1}$.
Through the properties of the inverse of a block matrix and the Woodbury matrix identity we find that
\begin{equation}\label{SMAI1}
	\begin{split}
		\C_{-xx} &= \left(\X^\top\R^{-1}\X - \X^\top\R^{-1}\Z\C_{zz}^{-1}\Z^\top\R^{-1}\X\right)^{-1} \\
		&= \left[\X^\top\left(\R^{-1}-\R^{-1}\Z\C_{zz}^{-1}\Z^\top\R^{-1}\right)\X\right]^{-1} \\
		&= \left(\X^\top\Vmat^{-1}\X\right)^{-1} \\
		\C_{-xz} &= -\C_{-xx}\X^\top\R^{-1}\Z\C_{zz}^{-1} \\
		\C_{-zx} &= -\C_{zz}^{-1}\Z^\top\R^{-1}\X\C_{-xx} \\
		\C_{-zz} &= \C_{zz}^{-1} + \C_{zz}^{-1}\Z^\top\R^{-1}\X\C_{-xx}\X^\top\R^{-1}\Z\C_{zz}^{-1} \text{.}
	\end{split}
\end{equation}
Now expand $\Wmat\C^{-1}\Wmat^\top$ and substitute (\ref{SMAI1}) into the expansion:
\begin{equation}\label{SMAI2}
	\begin{split}
		\Wmat\C^{-1}\Wmat^\top &= \X\C_{-xx}\X^\top + \X\C_{-xz}\Z^\top + \Z\C_{-zx}\X^\top + \Z\C_{-zz}\Z^\top \\
		&= \X\C_{-xx}\X^\top - \X\C_{-xx}\X^\top\R^{-1}\Z\C_{zz}^{-1}\Z^\top - \Z\C_{zz}^{-1}\Z^\top\R^{-1}\X\C_{-xx}\X^\top \\
		&\quad + \Z\C_{zz}^{-1}\Z^\top + \Z\C_{zz}^{-1}\Z^\top\R^{-1}\X\C_{-xx}\X^\top\R^{-1}\Z\C_{zz}^{-1}\Z^\top \text{.}
	\end{split}
\end{equation}
Define $\boldsymbol{\Omega} = \Z\C_{zz}^{-1}\Z^\top\R^{-1}$, take the last line of (\ref{SMAI2}), factor out $\X\C_{-xx}\X^\top$, and substitute $\boldsymbol{\Omega}$ where possible:
\begin{equation}\label{SMAI3}
	\Wmat\C^{-1}\Wmat^\top = \Z\C_{zz}^{-1}\Z^\top + \left(\I - \boldsymbol{\Omega}\right)\X\C_{-xx}\X^\top\left(\I - \boldsymbol{\Omega}^\top\right) \text{,}
\end{equation}
from which it follows that
\begin{equation}\label{SMAI4}
	\R^{-1}\Wmat\C^{-1}\Wmat^\top\R^{-1} = \R^{-1}\Z\C_{zz}^{-1}\Z^\top\R^{-1} + \R^{-1}\left(\I - \boldsymbol{\Omega}\right)\X\C_{-xx}\X^\top\left(\I - \boldsymbol{\Omega}^\top\right)\R^{-1} \text{.}
\end{equation}
Using the Woodbury matrix identity we see that for the first term $\R^{-1}\Z\C_{zz}^{-1}\Z^\top\R^{-1}$:
\begin{equation}\label{SMAI5}
	\Vmat^{-1} = \R^{-1} - \R^{-1}\Z\C_{zz}^{-1}\Z^\top\R^{-1} \Rightarrow \R^{-1}\Z\C_{zz}^{-1}\Z^\top\R^{-1} = \R^{-1} - \Vmat^{-1} \text{.}
\end{equation}
For the second term we can expand $\R^{-1}\left(\I - \boldsymbol{\Omega}\right)\X$ and use the Woodbury matrix identity again:
\begin{equation}\label{SMAI6}
	\begin{split}
		\R^{-1}\left(\I - \boldsymbol{\Omega}\right)\X &= \left(\R^{-1} - \R^{-1}\boldsymbol{\Omega}\right)\X \\
		&= \left(\R^{-1} - \R^{-1}\Z\C_{zz}^{-1}\Z^\top\R^{-1}\right)\X \\
		&= \Vmat^{-1}\X\text{.}
	\end{split}
\end{equation}
Similarly, $\X^\top\left(\I - \boldsymbol{\Omega}^\top\right)\R^{-1} = \X^\top\Vmat^{-1}$.
Substituting the results from (\ref{SMAI5}), (\ref{SMAI6}), and the expression for $\C_{-xx}$ from (\ref{SMAI1}) into (\ref{SMAI4}), we get:
\begin{equation}\label{SMAI7}
	\R^{-1}\Wmat\C^{-1}\Wmat^\top\R^{-1} = \R^{-1} - \Vmat^{-1} + \Vmat^{-1}\X\left(\X^\top\Vmat^{-1}\X\right)^{-1}\X^\top\Vmat^{-1} \text{.}
\end{equation}
Finally, we can substitute the result from (\ref{SMAI7}) into (\ref{SMAIstart}):
\begin{equation*}
	\begin{split}
		\R^{-1} - \R^{-1}\Wmat\C^{-1}\Wmat^\top\R^{-1} &= \R^{-1} - \left[\R^{-1} - \Vmat^{-1} + \Vmat^{-1}\X\left(\X^\top\Vmat^{-1}\X\right)^{-1}\X^\top\Vmat^{-1}\right] \\
		&= \R^{-1} - \R^{-1} + \Vmat^{-1} - \Vmat^{-1}\X\left(\X^\top\Vmat^{-1}\X\right)^{-1}\X^\top\Vmat^{-1} \\
		&= \Vmat^{-1} - \Vmat^{-1}\X\left(\X^\top\Vmat^{-1}\X\right)^{-1}\X^\top\Vmat^{-1} \text{,}
	\end{split}
\end{equation*}
which is, indeed, equal to $\Pmat$.

\section{Log-likelihood}\label{SM:LL}
Equation (\ref{eq:yPy}) states that for the quadratic term in the log-likelihood:
\begin{equation}\label{SM:LL:1}
	\y^\top \Pmat \y = \y^\top \R^{-1} \y - \y^\top \R^{-1} \Wmat_{d} \hat{\solution}_d - \y^\top \R^{-1} \Wmat_{s} \hat{\solution}_s \text{.}
\end{equation}
To see why (\ref{SM:LL:1}) is valid, we rewrite the right-hand side:
\begin{equation}\label{SM:LL;2}
	\begin{split}
		\y^\top \Pmat \y &= \y^\top \R^{-1} \y - \y^\top \R^{-1} \Wmat_{d} \hat{\solution}_d - \y^\top \R^{-1} \Wmat_{s} \hat{\solution}_s \\
		&= \y^\top \R^{-1} \y - \left(\y^\top \R^{-1} \Wmat_{d} \hat{\solution}_d + \y^\top \R^{-1} \Wmat_{s} \hat{\solution}_s\right) \\
		&= \y^\top \R^{-1} \y - \y^\top \R^{-1} \left(\Wmat_{d} \hat{\solution}_d + \Wmat_{s} \hat{\solution}_s\right) \\
		&= \y^\top \R^{-1} \y - \y^\top \R^{-1} \Wmat \hat{\solution} \text{,}
	\end{split}
\end{equation}
where $\Wmat = \left[\Wmat_d\;\Wmat_s\right]$ and $\hat{\solution} = \left[\hat{\solution_d}^\top\;\hat{\solution_s}^\top\right]^\top$.
Note that
\begin{equation*}
	\hat{\solution} = \C^{-1}\Wmat^\top\R^{-1}\y
\end{equation*}
which, when substituted into (\ref{SM:LL;2}), gives
\begin{equation*}
	\begin{split}
		\y^\top \Pmat \y &= \y^\top \R^{-1} \y - \y^\top \R^{-1} \Wmat \C^{-1}\Wmat^\top\R^{-1}\y \\
		&= \y^\top \left(\R^{-1} - \R^{-1} \Wmat \C^{-1}\Wmat^\top\R^{-1}\right)\y \text{,}
	\end{split}
\end{equation*}
implying that $\Pmat = \R^{-1} - \R^{-1} \Wmat \C^{-1}\Wmat^\top\R^{-1}$, which, as shown in Section \ref{SM:P}, is correct.

\section{AI matrix}\label{SM:AI}
We require that
\begin{equation}\label{SM:AI:1}
	\WV^\top \R^{-1} \WV - \WV^\top \R^{-1} \Wmat_d \mathbf{T}_d - \WV^\top \R^{-1} \Wmat_s \mathbf{T}_s = \WV^\top \Pmat \WV\text{.}
\end{equation}
The process of showing that these are equal is practically identical to the result provided in Section \ref{SM:LL} for the quadratic term of the log-likelihood, replacing $\y$ by $\WV$ throughout.
First note that the left-hand side of (\ref{SM:AI:1}) can be rewritten so that
\begin{equation}\label{SM:AI:2}
	\begin{split}
		\WV^\top \Pmat \WV &= \WV^\top \R^{-1} \WV - \left(\WV^\top \R^{-1} \Wmat_d \mathbf{T}_d + \WV^\top \R^{-1} \Wmat_s \mathbf{T}_s\right) \\
		&= \WV^\top \R^{-1} \WV - \WV^\top \R^{-1}\left(\Wmat_d \mathbf{T}_d + \Wmat_s \mathbf{T}_s\right) \\
		&= \WV^\top \R^{-1} \WV - \WV^\top \R^{-1} \Wmat \mathbf{T} \text{,}
	\end{split}
\end{equation}
where $\Wmat = \left[\Wmat_d\;\Wmat_s\right]$ and $\mathbf{T} = \left[\mathbf{T}_d^\top\;\mathbf{T}_s^\top\right]^\top$.
Then note that
\begin{equation*}
	\mathbf{T} = \C^{-1}\Wmat^\top\R^{-1}\WV \text{,}
\end{equation*}
which, when substituted into the right-hand side of (\ref{SM:AI:2}) gives
\begin{equation*}
	\begin{split}
		\WV^\top \Pmat \WV &= \WV^\top \R^{-1} \WV - \WV^\top \R^{-1} \Wmat \C^{-1}\Wmat^\top\R^{-1}\WV \\
		&= \WV^\top\left(\R^{-1} - \R^{-1} \Wmat \C^{-1}\Wmat^\top\R^{-1}\right)\WV \text{,}
	\end{split}
\end{equation*}
again correctly implying that $\Pmat = \R^{-1} - \R^{-1} \Wmat \C^{-1}\Wmat^\top\R^{-1}$.

\section{Residual scores}\label{SM:resscores}
To see that $\hat{\e}^\top\R^{-1}\left(\partial\R/\partial\phi_b\right)\R^{-1}\hat{\e} = \y^\top\Pmat\wv_b$.
Recall that
\begin{equation*}
	\Pmat = \R^{-1} - \R^{-1}\Wmat\C^{-1}\Wmat^\top\R^{-1} \text{,}
\end{equation*}
as shown in Section \ref{SM:P}.
By multiplying $\Pmat$ with $\y$ on the right we obtain:
\begin{equation*}
	\begin{split}
		\Pmat\y &= \R^{-1}\y - \R^{-1}\Wmat\C^{-1}\Wmat^\top\R^{-1}\y \\
		&= \R^{-1}\left(\y - \Wmat\C^{-1}\Wmat^\top\R^{-1}\y\right)
	\end{split} \text{.}
\end{equation*}
Note that $\Wmat\C^{-1}\Wmat^\top\R^{-1}\y = \Wmat\hat{\solution}$ which are the fitted values $\hat{\y}$.
We thus have
\begin{equation}\label{SM:resscores1}
	\begin{split}
		\Pmat\y &= \R^{-1}\left(\y - \hat{\y}\right) \\
		&= \R^{-1}\hat{\e}
	\end{split} \text{.}
\end{equation}
Now substitute (\ref{SM:resscores1}) and the definition of the residual working variates $\wv_b = \left(\partial\R/\partial\phi_b\right)\R^{-1}\hat{\e}$ into $\y^\top\Pmat\wv_b$:
\begin{equation*}
	\begin{split}
		\y^\top\Pmat\wv_b &= \left(\R^{-1}\hat{\e}\right)^\top\dfrac{\partial\R}{\partial\phi_b}\R^{-1}\hat{\e} \\
		&= \hat{\e}^\top\R^{-1}\dfrac{\partial\R}{\partial\phi_b}\R^{-1}\hat{\e}\text{,}
	\end{split}
\end{equation*}
as required.

\end{document}